\documentclass[acmsmall,screen,nonacm]{acmart}

\usepackage{amsmath}

\usepackage{graphicx}
\usepackage{booktabs}
\usepackage{array}
\usepackage{tabularx}
\usepackage{multirow}
\usepackage{enumitem}

\usepackage{textcomp}
\usepackage[table]{xcolor}
\usepackage{xspace}

\usepackage{algorithm}
\usepackage{algpseudocode}

\usepackage[most]{tcolorbox}

\usepackage{pifont}

\algnewcommand{\AlgHeader}[1]{%
  \Statex \textbf{#1}%
}

\algnewcommand{\AlgPhase}[1]{%
  \Statex
  \vspace{0.3em}
  \textbf{#1}%
}

\newcommand{\approach}{Rivet\xspace}

\newtcolorbox{answerbox}{
  enhanced,
  colback=gray!8,
  colframe=gray!70!black,
  boxrule=2pt,
  arc=4pt,
  boxsep=0pt,
  left=1.4em,
  right=1.4em,
  top=0.9em,
  bottom=0.9em,
  width=\linewidth,
  before skip=0.8em,
  after skip=0.8em,
  fontupper=\small
}

\def\BibTeX{%
  {\rm B\kern-.05em%
  {\sc i\kern-.025em b}\kern-.08em%
  T\kern-.1667em%
  \lower.7ex\hbox{E}\kern-.125emX}%
}

\begin{document}

\title{Rivet: Utility-Preserving Context Control for Repository-Level Coding Agents}


\author{Ruwei Pan}
\authornote{Both authors contributed equally to this research.}
\email{panruwei666@gmail.com}
\affiliation{%
  \institution{Peking University}
  \city{Beijing}
  \country{China}
}

\author{Linhao Wu}
\authornotemark[1]
\email{wulinhao26@stu.pku.edu.cn}
\affiliation{%
  \institution{Peking University}
  \city{Beijing}
  \country{China}
}

\author{Yakun Zhang}
\email{zhangyk@hit.edu.cn}
\affiliation{%
  \institution{Harbin Institute of Technology (Shenzhen)}
  \city{Shenzhen}
  \country{China}
}

\author{Zixiong Yang}
\email{zixiongyang02@gmail.com}
\affiliation{%
  \institution{Peking University}
  \city{Beijing}
  \country{China}
}

\author{Lu Zhang}
\email{zhanglucs@pku.edu.cn}
\affiliation{%
  \institution{Peking University}
  \city{Beijing}
  \country{China}
}

\author{Hongyu Zhang}
\authornote{Corresponding author.}
\email{hongyujohn@gmail.com}
\affiliation{%
  \institution{Chongqing University}
  \city{Chongqing}
  \country{China}
}

\renewcommand{\shortauthors}{Pan et al.}


\begin{abstract}

Repository-level code generation asks an agent to implement a task requirement inside an existing repository. To produce code that works in the repository, the agent needs \emph{repository evidence}, such as target locations, helper definitions, caller usages, and configuration constraints, which determine how the code should be located, integrated, and executed. Because the context budget cannot hold the entire repository, existing retrieval, compression, and structure-aware methods select context by task similarity, context length, or structural proximity, but these criteria do not directly measure whether the selected context contains this evidence. As a result, the selected context can appear relevant to the task yet miss the evidence that the implementation depends on. We call this failure mode \emph{context utility dilution}. In this paper, we propose \approach, a context control framework that preserves repository evidence for coding agents under a fixed context budget. \approach first uses Signal Grounding to select task-related code anchors by fusing lexical, semantic, and structural signals. Utility Linking then expands these anchors through dependency, usage, and configuration relations to collect repository evidence. Finally, Budget Shaping packs the evidence by priority and, when the budget is tight, reduces snippets to signatures or summaries before dropping them. We evaluate \approach on two executable repository-level benchmarks, DevEval and RepoExec, with four backbone models, two context budgets (4K and 8K tokens), and eight baselines. \approach achieves the highest Pass@1 in all 16 settings and improves over the strongest baseline by 3.73--5.15 percentage points on average. Moreover, GPT-5-mini with \approach outperforms GPT-5 without \approach on both benchmarks while reducing inference cost by 73.3\%--84.5\%.

\end{abstract}

\ccsdesc[500]{Software and its engineering~Automatic programming}

\keywords{Code Generation, Large Language Models}

\maketitle


\section{Introduction}

Given a task requirement and an existing repository, repository-level code generation asks an agent to produce code that satisfies the requirement within that repository~\cite{yang2024swe, liu2024repobench, zhang2024autocoderover,bairi2024codeplan}.
Unlike isolated programming tasks, the produced code must work with existing files, definitions, dependencies, and configurations in the target repository~\cite{jimenez2024swe}.
To produce code that works in the repository, the agent needs \emph{repository evidence}, i.e., repository information that constrains the implementation, such as target locations, helper definitions, caller usages, and configuration constraints.
Repository context is therefore useful only to the extent that it contains such evidence.

However, a repository rarely fits into the context budget of a coding agent, so the agent can only see a selected part of it~\cite{shi2025longcodezip}.
Existing context-control methods address this constraint through retrieval-based selection, compression-based shortening, and structure-aware selection~\cite{lewis2020retrieval,liu2024repobench,pan2024llmlingua,hu2026line,le2026not,wu2024repoformer,wang2025rlcoder}.
Retrieval-based methods select code snippets according to their similarity to the task requirement~\cite{lewis2020retrieval,liu2024repobench}.
Compression methods shorten retrieved repository context or agent observations to fit the context budget~\cite{pan2024llmlingua}.
Structure-aware methods use code structure, such as dependencies or call relations, to retrieve context connected to the task~\cite{hu2026line,le2026not,cheng2024dataflow}.
These criteria provide useful signals for context construction, but a high relevance score, a compact representation, or structural proximity alone does not ensure that the context includes the helper definitions, caller usages, and configuration constraints needed for a particular implementation.
\textbf{We call the resulting failure mode \emph{context utility dilution}: the selected context appears relevant to the task but lacks the evidence needed to implement the required code correctly in the repository (Section~\ref{sec:motivating}).}

In this paper, we propose \textbf{\approach}, a context control framework that preserves repository evidence for repository-level coding agents.
Our key observation is that much of this evidence is reachable from the code that the task touches, through static relations such as imports, symbol references, call sites, and configuration links.
Therefore, instead of ranking snippets independently by their similarity to the task, \approach first locates where the task lands in the repository and then follows these relations to collect evidence.
Following this observation, \approach constructs repository context in three stages.
\ding{182} \emph{Signal Grounding} first uses lexical, semantic, and structural signals to identify task-related code anchors that indicate where the required code is likely to be implemented or what existing snippets it may depend on.
\ding{183} \emph{Utility Linking} then expands these anchors through dependency, usage, and configuration relations to collect helper definitions, caller usages, and configuration constraints.
\ding{184} \emph{Budget Shaping} finally packs the evidence by priority tiers and, when the budget is tight, reduces snippets to signatures or summaries before dropping them.
In this way, \approach spends the budget on evidence that constrains the implementation rather than on text that merely resembles the task.

We evaluate \approach on two executable repository-level coding benchmarks, \textbf{DevEval}~\cite{li2024deveval} and \textbf{RepoExec}~\cite{le2025impacts}.
Across four backbone models and two context budgets, 4K and 8K tokens, we compare \approach against retrieval, compression, and structure-aware baselines under the same agent scaffold.
\approach achieves the highest Pass@1 in all 16 settings and improves over the strongest baseline by 3.73--5.15 percentage points on average.
With a 4K budget, it outperforms the strongest baseline given an 8K budget on DevEval for all four backbone models, and GPT-5-mini with \approach outperforms GPT-5 without \approach while reducing inference cost by 73.3\%--84.5\%.
Ablation studies further show that all three stages contribute, with Signal Grounding and Utility Linking accounting for the largest gains.

In summary, this paper makes the following contributions:

\setlist[itemize]{leftmargin=1em, labelsep=0.6em, itemsep=2pt}

\begin{itemize}

\item \textbf{We identify and characterize \emph{context utility dilution}, a failure mode of context construction for repository-level coding agents.}
This failure mode occurs when the selected context appears relevant to the task but lacks the repository evidence needed to locate, integrate, and execute the required code in the target repository.
We illustrate this failure mode with a motivating example (Section~\ref{sec:motivating}) and a qualitative case study (Figure~\ref{fig:case}).

\item \textbf{We propose \approach, a context control framework that preserves repository evidence under constrained context budgets.}
\approach locates task-related code anchors, follows static repository relations from them to collect dependency, usage, and configuration evidence, and packs this evidence under the budget with a representation ladder.
This design helps coding agents keep target locations, helper definitions, caller usages, and configuration constraints within the context budget.

\item \textbf{We evaluate \approach on two repository-level coding benchmarks, DevEval and RepoExec, with four backbone models and two context budgets, 4K and 8K tokens.}
The results show that \approach achieves an average Pass@1 of 55.74 across all settings and consistently outperforms retrieval, compression, and structure-aware baselines under the same agent scaffold.
With half the context budget or a cheaper backbone, \approach still matches or exceeds stronger configurations at a fraction of the inference cost.

\end{itemize}

\section{Motivating Example}
\label{sec:motivating}

\begin{figure}[t]
\centering
\includegraphics[width=0.55\linewidth]{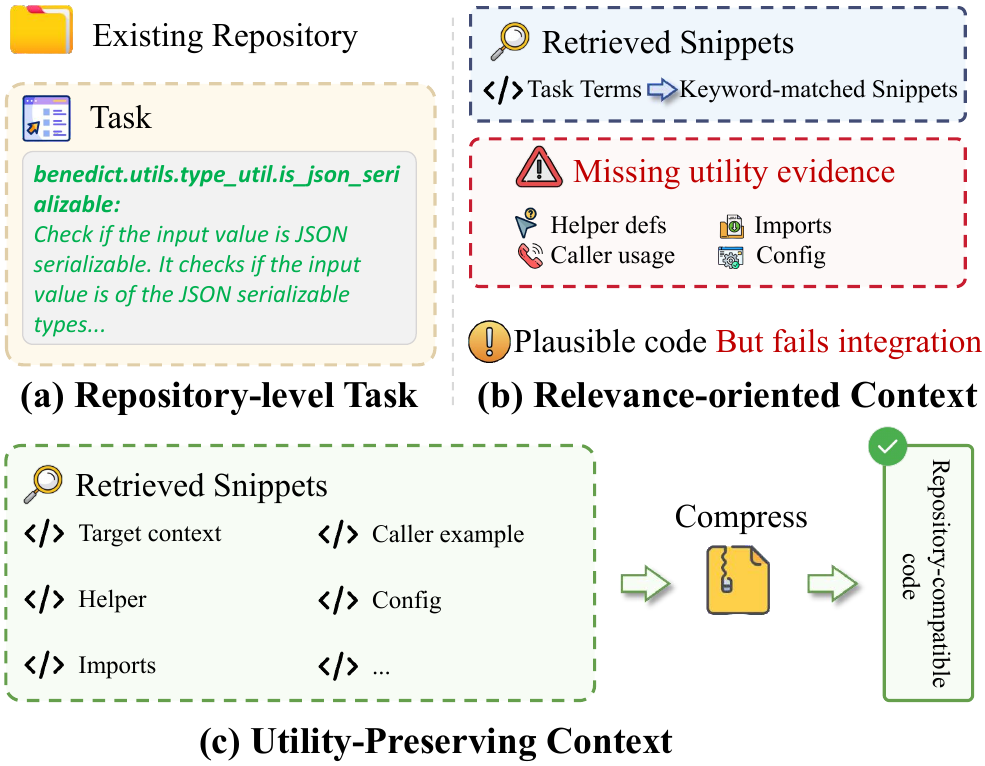}
\caption{Motivating example of context utility dilution. (b) Relevance-oriented retrieval returns keyword-matched snippets but misses helper definitions, caller usages, imports, and configuration. (c) Utility-preserving context retains this evidence under the same budget.}
\vspace{-2mm}
\label{fig:motivating}
\end{figure}

Figure~\ref{fig:motivating} shows a DevEval task in which an agent needs to implement the utility function \path{is_json_serializable} in the module
\path{benedict.utils.type_util} of an existing Python repository.
The target function is expected to determine whether a given value is JSON serializable.
Although the requirement is concise, the required code cannot be derived from the requirement alone~\cite{ding2023crosscodeeval}.
To produce code that works in this repository, the agent needs repository evidence, including the target location, relevant helper definitions, caller usages, and configuration constraints.

A relevance-oriented retrieval method may return code snippets related to JSON serialization or type checking.
These snippets appear relevant because they share terms with the requirement, but they may miss the evidence that constrains how the function should be implemented in this repository.
For example, without helper definitions, the agent may reimplement behavior that should reuse project-specific utilities. Without caller usages, it may produce an interface or return value that does not match how the function is used. Without configuration constraints, it may generate code that cannot execute correctly in the project environment~\cite{wang2023automatically}.
Under a limited context budget, these relevant but incomplete snippets can occupy the context and leave little room for the evidence that actually constrains the implementation~\cite{gu2025retrieve}.
For this task, useful context should include not only snippets related to JSON serialization, but also the target module \texttt{benedict.utils.type\_util}, relevant helper definitions, caller usages, dependencies, and configuration constraints.
With such evidence, the agent can determine not only what behavior to implement, but also how the required code should align with the existing repository.

This example illustrates \emph{context utility dilution}: within a fixed budget, snippets that merely resemble the task dilute the context, so the evidence that determines how \texttt{is\_json\_serializable} fits into the repository is either missing or crowded out.
It suggests that context construction should start from where the task lands in the repository and collect the evidence connected to it, rather than rank snippets by similarity alone.
This observation motivates the design of \approach (Section~\ref{sec:approach}).

\section{Approach}
\label{sec:approach}

\begin{figure*}[t]
\centering
\includegraphics[width=0.85\linewidth]{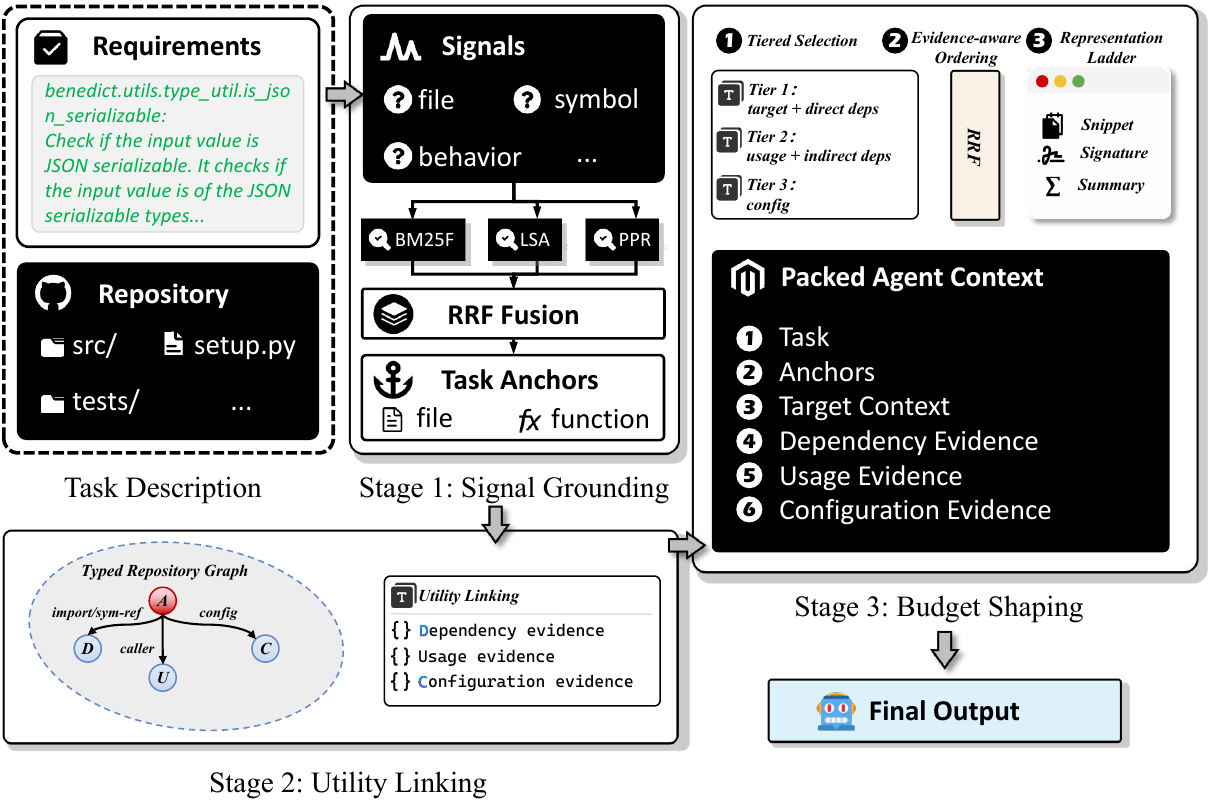}
\caption{
Overview of \approach. Signal Grounding fuses BM25F, LSA, and PPR rankings to select task anchors; Utility Linking expands the anchors along import/symbol-reference, caller, and configuration edges; Budget Shaping packs the collected evidence by priority tier with a representation ladder.
}

\label{fig:overview}
\end{figure*}

Given a task requirement, a repository, and a context budget, \approach serves as a context-construction component for a repository-level coding agent.
It constructs the repository context that is injected into the agent before code generation.
The goal is to preserve the repository evidence that the implementation depends on, including target-location information, helper definitions, caller usages, and configuration constraints.
\approach operates at two granularities: Signal Grounding ranks repository files to locate where the task lands, while Utility Linking and Budget Shaping operate on code snippets within and around the selected files.
\textbf{Signal Grounding} identifies task-related code anchors using lexical matching, semantic similarity, and structural propagation signals.
\textbf{Utility Linking} expands these anchors through dependency, usage, and configuration relations to collect the repository evidence needed by the required code.
\textbf{Budget Shaping} retains the resulting evidence under the context budget by prioritizing essential evidence and using compact representations when necessary.
Figure~\ref{fig:overview} summarizes the pipeline. We next introduce the problem setting, the repository evidence graph, the query construction, and the three stages of \approach.

\subsection{Problem Formulation}
\label{sec:overview}

\textbf{Problem setting.}
A task is specified by a natural-language requirement $q$.
A repository consists of a set of files $F$ and is parsed into a set of code snippets $U=\{u_1,\dots,u_N\}$, including functions, methods, classes, import blocks, and configuration blocks, each of which belongs to one file in $F$.
Given $q$, $U$, and a context budget $B$, \approach selects and renders a repository context $C$ such that $\mathrm{tok}(C)\le B$.
The target location is not given to \approach during context construction.
Instead, \approach identifies task-related code anchors from the repository through Signal Grounding.
When the requirement explicitly mentions code entities, such as the qualified name of the target function, \approach exploits these mentions through query construction and a code-mention prior, but it never receives the ground-truth location.

\textbf{Repository evidence graph.}
We represent the repository as a directed, typed, weighted graph $G=(V,E,\tau,\omega)$ with $V=U$.
Each node is a code snippet, and each edge $e=(u\!\to\!v)$ represents a repository relation with type $\tau(e)$ and weight $\omega(e)$.
We consider four types of relations that expose repository evidence around candidate code anchors.
\textsc{import} and \textsc{symbol-ref} edges connect a snippet to definitions, imports, and referenced symbols that it depends on, thereby exposing helper definitions and dependency evidence.
\textsc{caller} edges connect a snippet to existing call sites or usages that indicate how it is invoked.
\textsc{config} edges associate configuration or packaging information with relevant repository code~\cite{peng2024less}.
The edge weight $\omega(e)$ reflects how strongly the relation indicates evidence needed by the required code.
All three stages of \approach operate on this graph.

\textbf{Query construction.}
Before Signal Grounding, \approach converts the natural-language requirement $q$ into a weighted query term set $Q=\{(w,\gamma_w)\}$.
The terms are extracted from the requirement text and any explicit code-like mentions in $q$, such as identifiers, file paths, function signatures, and package names, and are split on camel-case and snake-case boundaries.
Terms from explicit code-like mentions receive higher weights than general natural-language terms.
The same query terms are used for lexical and semantic scoring, so that both signals are computed from the same task representation.

\subsection{Signal Grounding}
\label{sec:anchor}

Signal Grounding aims to identify task-related anchor files from the repository.
In this stage, $u$ denotes a repository file, and the structural signal runs on the file-level import graph $G_F$, obtained by projecting the \textsc{import} edges of $G$ onto files, i.e., file $u$ links to file $v$ if a snippet in $u$ imports a definition from $v$.
These anchors serve as candidate locations for the required code and as starting points for collecting repository evidence in Utility Linking.
To this end, \approach combines three complementary grounding signals: lexical matching, semantic similarity, and structural propagation.
The lexical signal captures explicit code-like mentions in the requirement, such as identifiers, file paths, and package names.
The semantic signal captures files that are related to the task in meaning, even when their surface terms differ from the requirement.
The structural signal propagates relevance over $G_F$ to find files connected to high-relevance files through import relations.
These signals address different failure cases of anchor localization. Lexical matching can miss renamed or project-specific helpers, semantic similarity can prefer textually related but unnecessary files, and structural propagation depends on reliable task-relevant seeds.
Therefore, \approach fuses the rankings from the three signals to select the final task-related code anchors.

\textbf{Lexical signal.}
The lexical signal identifies files that explicitly match code-like mentions in the requirement, such as identifiers, file paths, function signatures, or package names.
We score each repository file $u$ with a fielded BM25~\cite{robertson2009probabilistic} model over four fields
$\mathcal{F}=\{\textit{path},\textit{symbol},\textit{sigdoc},\textit{content}\}$,
which respectively capture the file path, defined symbol names, signature and docstring lines, and the identifier stream.
Field weighting and length normalization reduce the bias toward long files that contain many incidental term matches.
The resulting lexical score is denoted as $s_{\text{lex}}(u)$.

\textbf{Semantic signal.}
The semantic signal identifies files that are related to the task in a latent term space, even when their surface terms differ.
For example, a requirement may describe JSON serialization, while a reusable helper may use a project-specific name.
To capture such cases without training a repository-specific retriever, we build a TF-IDF term--file matrix over file content and documentation and apply latent semantic analysis~\cite{deerwester1990indexing}.
Each file and the task requirement are projected into the same latent space, and their cosine similarity is used as the semantic score $s_{\text{sem}}(u)$.
For repositories too small to admit a stable factorization, we fall back to TF-IDF cosine similarity in the original term space.

\textbf{Structural signal.}
The structural signal identifies files that may be needed by the required code even when they are not textually similar to the requirement.
We use the normalized sum of the lexical and semantic scores as seed relevance scores and diffuse them over the file-level import graph $G_F$ with Personalized PageRank~\cite{haveliwala2002topic}.
This propagation allows a file connected to a high-relevance file to inherit relevance even if it is lexically invisible to the requirement.
The resulting structural score is denoted as $s_{\text{str}}(u)$.

\textbf{Anchor selection.}
The three signals induce rankings $\mathrm{rank}_c(u)$ for
$c\in\{\text{lex},\text{sem},\text{str}\}$.
We fuse these rankings with reciprocal-rank fusion~\cite{cormack2009reciprocal}:
\begin{equation}
\label{eq:rrf}
\rho_0(u)=
\sum_{c\in\{\text{lex},\text{sem},\text{str}\}}
\frac{1}{\kappa+\mathrm{rank}_c(u)} .
\end{equation}
This fusion combines complementary grounding signals without requiring their scores to share the same scale.
We further add a deterministic prior $\pi(u)$ for files that match code-like mentions explicitly appearing in the requirement, such as identifiers, file paths, function signatures, or package names.
The final grounding score is \(\rho(u)=\rho_0(u)+\pi(u)\), which is used for anchor selection and subsequent evidence propagation.
\approach keeps the top-$k$ files as the task anchor set $A$, and the code snippets in these files serve as the starting points of Utility Linking.
These task anchors are then passed to Utility Linking, which expands them into the repository evidence needed by the required code.

\subsection{Utility Linking}
\label{sec:closure}

Signal Grounding identifies task-related code anchors, but anchors alone do not provide all repository evidence needed by the required code.
Utility Linking therefore expands from these anchors to collect potentially useful repository evidence through static relations.
It performs this expansion in two steps.
First, starting from the anchor set $A$, \approach traverses the repository evidence graph $G$ and gathers snippets connected to the anchors through dependency, usage, and configuration relations.
Second, \approach assigns each gathered evidence snippet a relevance score that reflects how closely it is connected to the required code.

\textbf{Evidence categories.}
\approach collects three categories of repository evidence around task-related code anchors.
\ding{182} Dependency evidence includes definitions, imports, and referenced symbols reached from an anchor through \textsc{import} and \textsc{symbol-ref} edges.
This evidence helps the agent reuse project-specific definitions and resolve required dependencies.
\ding{183} Usage evidence includes existing call sites or usages recovered from \textsc{caller} edges.
This evidence indicates how an anchor or related snippet is invoked in the repository.
\ding{184} Configuration evidence includes packaging, runtime, and configuration snippets connected through \textsc{config} edges.
This evidence constrains how the required code should be integrated and executed in the target repository.
Together with the task-related code anchors, these categories form the candidate repository evidence set used for context construction.
Each evidence snippet is extracted at snippet granularity, such as a definition block with its signature and a bounded context, rather than as a whole file.

\textbf{Evidence relevance.}
The collected evidence induces an evidence graph $G_E$, whose nodes are the task-related code anchors and the gathered evidence snippets, and whose edges inherit their relation types from $G$.
Rather than assigning fixed scores to different evidence categories, \approach reuses the propagation mechanism in the Signal Grounding stage on $G_E$.
The propagation is seeded by the anchor relevance and diffused with type weights, e.g.,
$\omega_{\textsc{import}}>\omega_{\textsc{symbol-ref}}>\omega_{\textsc{caller}}>\omega_{\textsc{config}}$.
The resulting score $\psi(z)$ measures how strongly an evidence snippet $z$ is connected to the required code.
Evidence directly connected to an anchor receives a higher score, while incidentally connected evidence receives a lower score.
This score is passed to Budget Shaping, which decides how to retain the collected evidence under the context budget.

\subsection{Budget Shaping}
\label{sec:packaging}

Budget Shaping converts the collected repository evidence into a packed repository context under the context budget $B$.
Its goal is to retain the evidence needed by the required code, while using compact representations when the full evidence set exceeds the budget.
Budget Shaping makes three decisions: which priority tier each evidence snippet belongs to, how compactly each snippet should be represented, and in what order snippets should be considered for the budget.

\textbf{Evidence tiers and representation ladder.}
Each evidence snippet is assigned a priority tier according to its role in supporting the required code.
Task-related code anchors and their direct dependencies are placed in the highest tier.
Usage evidence and indirect dependencies are placed in the next tier.
Configuration evidence is placed in the last tier. These tiers provide a heuristic ordering for budget allocation rather than a task-independent measure of evidence importance.
Within each tier, an evidence snippet can be rendered at different levels of detail through a representation ladder
$\langle\textit{full snippet},\textit{signature},\textit{compact summary}\rangle$.
When the budget is tight, \approach reduces the representation detail of a selected snippet before dropping it.
For high-priority evidence, \approach keeps at least its signature whenever possible.
This design keeps more types of repository evidence under constrained budgets, instead of spending the budget on a few long snippets.

\textbf{Evidence-aware ordering.}
Evidence snippets are considered tier by tier.
Within each tier, \approach orders snippets by combining two rankings.
The first ranking is the anchor relevance $\mathrm{rel}(z)=\rho(a_z)$, i.e., the grounding score of the anchor $a_z$ that introduces evidence snippet $z$.
The second ranking is the evidence relevance $\psi(z)$, which reflects how strongly $z$ is connected to the required code in the evidence graph.
We combine these rankings using reciprocal-rank fusion:
\begin{equation}
\label{eq:pack}
\mathrm{key}(z)=
\bigg(
\mathrm{tier}(z),
-\!\!\!\sum_{c\in\{\text{rel},\psi\}}\!\!\!
\frac{1}{\kappa+\mathrm{rank}_c(z)}
\bigg).
\end{equation}
The tier is used as the primary key, so that evidence directly needed by the required code is considered first.
The fused ranking is used as the secondary key, so that the budget is spent on evidence that is both task-relevant and strongly connected to the required code.
The selected and rendered snippets, together with the task requirement and task-related code anchors, form the packed repository context $C$ injected into the coding agent.

\begin{algorithm}[t]
\caption{Repository context construction in \approach}
\label{alg:rivet}
\begin{algorithmic}[1]
\Require requirement $q$, repository files $F$ with code snippets $U$, budget $B$
\Ensure repository context $C$ with $\mathrm{tok}(C)\le B$

\State $Q \gets \textsc{ConstructQuery}(q)$
\State $G \gets \textsc{BuildEvidenceGraph}(U)$
\State $G_F \gets \textsc{ProjectImports}(G,F)$ \Comment{file-level import graph}

\For{each file $f \in F$}
    \State $s_{\mathrm{lex}}(f)\gets\textsc{LexicalScore}(Q,f)$
    \State $s_{\mathrm{sem}}(f)\gets\textsc{SemanticScore}(Q,f)$
\EndFor

\State $s_{\mathrm{str}}\gets\textsc{Propagate}(\textsc{Norm}(s_{\mathrm{lex}}+s_{\mathrm{sem}}),G_F,\alpha)$
\State $\pi_q \gets \textsc{CodeMentionPrior}(q,F)$
\State $A \gets \textsc{TopK}\big(\textsc{RRF}(s_{\mathrm{lex}},s_{\mathrm{sem}},s_{\mathrm{str}})+\pi_q,k\big)$

\State $\mathcal{E}\gets\textsc{LinkEvidence}(A,G)$ \Comment{dependency / usage / configuration}
\State $G_E \gets \textsc{InduceEvidenceGraph}(A,\mathcal{E},G)$
\State $\psi\gets\textsc{Propagate}(\rho|_A,G_E,\alpha)$

\State $C\gets\textsc{ShapeBudget}(q,A,\mathcal{E},\psi,B)$ \Comment{tiers + representation ladder}
\State \Return $C$
\end{algorithmic}

\end{algorithm}

\section{Experimental Setup}

We evaluate \approach by answering the following three research questions (RQs):

\textbf{RQ1: How effective is \approach under constrained context budgets?}
We compare \approach with retrieval, compression, and structure-aware baselines on DevEval and RepoExec.
This experiment evaluates whether preserving repository evidence improves repository-level code generation under the same backbone model, agent scaffold, and context budget.

\textbf{RQ2: How does \approach compare with larger context budgets and stronger backbone models in terms of accuracy and inference cost?}
We compare \approach with two common alternatives for improving accuracy, namely doubling the context budget and switching to a stronger backbone model under the same Mini-SWE-Agent scaffold, in terms of both Pass@1 and inference cost.

\textbf{RQ3: How do \approach's stages and evidence types contribute to its effectiveness?}
We conduct two levels of ablation studies. First, we disable Signal Grounding, Utility Linking, and Budget Shaping to evaluate the contribution of each stage. Second, we independently disable dependency-, usage-, and configuration-based expansion to examine the contribution of different repository relations within Utility Linking.

\subsection{Benchmarks}

Following prior work on repository-level code generation~\cite{hu2026line, le2026not}, we conduct our main experiments on two executable repository-level coding benchmarks, DevEval and RepoExec.
We choose them because they provide repository-level tasks with executable evaluation protocols, allowing us to measure whether the generated code works within the target repository.
\ding{182} \textbf{DevEval~\cite{li2024deveval}} is a repository-level code generation benchmark built from real-world repositories, containing 1,825 function-level generation tasks.
Each task provides a natural-language requirement and the corresponding original repository, and the model is expected to implement the required function in the target codebase.
We use DevEval as the main benchmark because its tasks require repository evidence such as target locations, project-specific definitions, dependencies, and interface constraints.
\ding{183} \textbf{RepoExec~\cite{le2025impacts}} is a repository-level code generation benchmark containing 355 function-level generation tasks with ground-truth annotations.
Compared with DevEval, RepoExec places stronger emphasis on whether generated code can execute correctly within the repository context.
We use RepoExec as an external validation benchmark to examine whether the gains of \approach transfer to different repositories and test suites.

During context construction, all methods use only the task requirement and the original repository files.
Reference implementations, reference dependencies, and benchmark test outcomes are reserved for evaluation and analysis, and are not used to construct the repository context.
This protocol avoids information leakage and ensures that all methods are compared under the same available information.
The only exception is Target-Local, which by design receives the benchmark-provided target location.

\subsection{Metrics}

We use \textbf{Pass@1~\cite{chen2021evaluating}} as the primary evaluation metric. It measures the percentage of tasks for which the generated code passes the official tests. Given $n$ tasks, Pass@1 is defined as:
\begin{equation}
\mathrm{Pass@1}
=
\frac{1}{n}
\sum_{i=1}^{n}
\mathbf{1}\left[\mathrm{Eval}(y_i)=\mathrm{pass}\right],
\end{equation}
where $y_i$ denotes the code generated for the $i$-th task, and $\mathrm{Eval}(\cdot)$ denotes the official benchmark evaluator. For repository-level code generation, Pass@1 directly reflects whether the generated code satisfies the task requirement within the target repository.
To test whether the gains are statistically robust, we compare \approach with the strongest non-\approach baseline using paired bootstrap resampling over tasks and estimate 95\% confidence intervals of the Pass@1 difference for each benchmark and context budget, pooling the four backbone models.

\subsection{Baselines}

We compare \approach with representative baselines for repository-level code generation, including \textbf{retrieval-based, compression-based, and structure-aware methods}.
Target-Local is included as a special local-context baseline because it uses the benchmark-provided target location and evaluates whether local context around the target alone is sufficient.
The remaining baselines construct or reduce repository context from the task requirement and repository files under the same context budget.

\textit{\underline{Local-context baseline.}}
\textbf{Target-Local} provides the task requirement together with a fixed local context around the benchmark-provided target location, including surrounding code from the target file and entry-point hints.
Unlike the other baselines, it does not evaluate automatic context discovery.
Instead, it serves as a reference with access to the ground-truth target location, testing whether target-location context alone is sufficient.

\textit{\underline{Retrieval-based baseline.}}
\textbf{RAG~\cite{lewis2020retrieval}} divides the repository into code snippets and retrieves snippets according to their similarity to the task requirement.
It keeps the most relevant snippets under the context budget and concatenates them as the repository context.

\textit{\underline{Compression-based baselines.}}
\ding{182} \textbf{LLMLingua2~\cite{pan2024llmlingua}} takes the retrieved repository context as input, estimates the importance of tokens or short spans with its pretrained compressor, and removes low-importance content until the context budget is met. The remaining text is used as the repository context. \ding{183} \textbf{LLM Summarize}, following the baseline setting in~\cite{wang2026swe}, asks an LLM to rewrite the retrieved repository context into a shorter summary under the context budget. The summary is then used as the repository context. \ding{184} \textbf{LongCodeZip~\cite{shi2025longcodezip}} ranks function-level code snippets by their relevance to the task requirement and further shortens the retained snippets by selecting important code blocks under the context budget. The shortened code context is used as the repository context. \ding{185} \textbf{SWE-Pruner~\cite{wang2026swe}} asks the agent to produce an explicit pruning goal, such as focusing on error handling or locating a relevant function. Guided by this goal, a lightweight neural skimmer selects relevant lines from raw agent observations, such as file-reading or search outputs. The remaining lines are used as the pruned agent observation.

\textit{\underline{Structure-aware baselines.}}
\ding{182} \textbf{InlineCoder~\cite{hu2026line}} performs repository-level code generation with call-graph context inlining. It first generates a draft completion as an anchor for the target function, and then enriches the prompt with upstream callers and downstream callees. \ding{183} \textbf{Hydra~\cite{le2026not}} treats code as structured programs rather than natural-language chunks, builds hierarchical indices over functions, classes, and variables, and combines dependency-aware retrieval with similarity-based retrieval to provide required dependencies and usage examples.

\subsection{Implementation Details}
\label{sec:implementation_details}

Following prior work~\cite{wang2026swe}, we use \textbf{Mini-SWE-Agent} as the unified agent scaffold for all methods.
We evaluate four backbone models: \textbf{DeepSeek-V3.2}, \textbf{GPT-5-mini}, \textbf{Qwen3.5-397B-A17B}, and \textbf{MiMo-V2.5}.
We use temperature 0 for all models that support it and the default decoding setting for GPT-5 models.
We set the context budgets to 4K and 8K tokens.
To ensure a fair comparison, all methods use the same task requirement, repository files, backbone model, prompt template, context budget, decoding strategy, agent scaffold, and execution environment.
The only difference is how each method constructs the repository context under the same budget.
The generated code is executed and evaluated in a consistent Python 3.9 environment.

The 4K and 8K budgets are chosen to evaluate repository context construction under practical cost constraints.
The 4K budget represents a tight setting, where the agent cannot include many repository files and must prioritize essential repository evidence.
The 8K budget provides a less restrictive setting for testing whether the benefit of \approach remains when more repository context can be included.
Together, these budgets allow us to examine whether \approach improves accuracy by preserving repository evidence rather than by simply increasing the amount of repository text.

\textbf{Hyperparameters of \approach.}
We use the same hyperparameter values for all benchmarks, backbone models, and context budgets.
In Signal Grounding, \approach keeps the top-$k{=}6$ files as anchors, a small set that leaves most of the budget for the evidence linked from them.
The fielded BM25 model uses the standard saturation parameter $k_1{=}1.2$~\cite{robertson2009probabilistic}, with field weights (and length-normalization parameters $b$) of 3.0 (0.40), 2.5 (0.50), 1.5 (0.60), and 1.0 (0.75) for the \textit{path}, \textit{symbol}, \textit{sigdoc}, and \textit{content} fields, so that the precise path and symbol fields dominate the noisier identifier stream.
For the same reason, query terms from explicit code mentions receive weights of 2.0 (symbols) and 1.5 (file names), compared with 1.0 for other terms.
Latent semantic analysis uses $\min(256, |F|{-}1)$ latent dimensions and falls back to TF-IDF cosine similarity for repositories with fewer than 20 files, where a truncated factorization is unstable.
Personalized PageRank runs for 20 iterations with damping factors $\alpha{=}0.6$ in Signal Grounding and $\alpha{=}0.7$ in Utility Linking; both values are lower than the conventional 0.85, which keeps the propagated relevance close to the task-related seeds.
Both reciprocal-rank fusions use $\kappa{=}60$, the value recommended by Cormack et al.~\cite{cormack2009reciprocal}, and the code-mention prior $\pi(u)$ adds 0.05 to the fused score of explicitly mentioned files, which places them ahead of the remaining candidates.
The edge weights follow the priority order of the evidence tiers: $\omega_{\textsc{import}}{=}1.0$ (0.7 for indirect imports), $\omega_{\textsc{symbol-ref}}{=}0.8$, $\omega_{\textsc{caller}}{=}0.5$, and $\omega_{\textsc{config}}{=}0.3$.
Utility Linking retains at most 12 dependency, 10 usage, and 10 configuration snippets, and Budget Shaping caps each section of the packed context, e.g., at 1,400, 420, 240, and 140 tokens for the target context and the dependency, usage, and configuration evidence under the 4K budget, and at 2,500, 1,200, 800, and 500 tokens under the 8K budget.

\section{Experimental Results}

\subsection{RQ1: Accuracy Comparison}

\definecolor{DeepBlue}{RGB}{77,88,234}
\definecolor{LightBlue}{RGB}{235,241,250}
\newcommand{\ourscell}[1]{\cellcolor{LightBlue}\textcolor{DeepBlue}{\textbf{#1}}}

\begin{table}[t]
\centering
\caption{Pass@1 comparison on DevEval and RepoExec under 4K and 8K context budgets.}
\label{tab:rq1_main_results}
\scriptsize
\setlength{\tabcolsep}{2.5pt}
\renewcommand{\arraystretch}{1.03}

\resizebox{0.9\columnwidth}{!}{
\begin{tabular}{
>{\centering\arraybackslash}p{3cm}
>{\raggedright\arraybackslash}p{1.85cm}
>{\centering\arraybackslash}p{0.72cm}
>{\centering\arraybackslash}p{0.72cm}
>{\centering\arraybackslash}p{0.72cm}
>{\centering\arraybackslash}p{0.72cm}
}
\toprule
\multirow{2}{*}{\textbf{Model}}
& \multirow{2}{*}{\textbf{Method}}
& \multicolumn{2}{c}{\textbf{DevEval}}
& \multicolumn{2}{c}{\textbf{RepoExec}} \\
\cmidrule(lr){3-4} \cmidrule(lr){5-6}
&
& \textbf{4K}
& \textbf{8K}
& \textbf{4K}
& \textbf{8K} \\
\midrule

\multirow{9}{*}{\textbf{DeepSeek-V3.2}}
& Target-Local  & 43.29 & 44.16 & 33.24 & 34.37 \\
& RAG           & 49.64 & 49.92 & 36.90 & 35.77 \\
& LLMLingua2    & 36.33 & 36.93 & 32.68 & 32.68 \\
& LLM Summarize & 30.90 & 30.58 & 28.73 & 27.89 \\
& LongCodeZip   & 46.19 & 46.96 & 38.87 & 36.62 \\
& SWE-Pruner    & 52.66 & 53.42 & 35.49 & 37.18 \\
& InlineCoder   & 51.12 & 54.74 & 38.87 & 42.54 \\
& Hydra         & 31.45 & 37.10    & 34.93 & 36.34 \\
& \ourscell{\approach} & \ourscell{59.89} & \ourscell{59.56} & \ourscell{49.01} & \ourscell{51.27} \\
\midrule

\multirow{9}{*}{\textbf{GPT-5-mini}}
& Target-Local  & 53.59 & 52.49 & 39.72 & 40.00 \\
& RAG           & 58.08 & 58.08 & 45.35 & 43.38 \\
& LLMLingua2    & 49.04 & 48.11 & 37.46 & 37.18 \\
& LLM Summarize & 41.04    & 43.01    & 49.01 & 49.30 \\
& LongCodeZip   & 55.34 & 55.95 & 45.35 & 47.61 \\
& SWE-Pruner    & 60.00 & 59.62 & 46.48 & 47.04 \\
& InlineCoder   & 60.27 & 64.38 & 44.79 & 49.58 \\
& Hydra         & 38.36 & 46.79    & 40.85 & 41.97 \\
& \ourscell{\approach} & \ourscell{64.93} & \ourscell{66.25} & \ourscell{49.30} & \ourscell{53.52} \\
\midrule

\multirow{9}{*}{\textbf{Qwen3.5-397B-A17B}}
& Target-Local  & 44.60 & 46.25 & 39.44 & 37.75 \\
& RAG           & 53.42 & 52.27 & 41.13 & 42.82 \\
& LLMLingua2    & 43.07 & 40.82 & 36.62 & 36.90 \\
& LLM Summarize & 36.05 & 35.84 & 31.83 & 33.80 \\
& LongCodeZip   & 50.85 & 49.26 & 47.61 & 45.07 \\
& SWE-Pruner    & 56.22 & 54.30 & 45.35 & 44.51 \\
& InlineCoder   & 47.84 & 57.53 & 45.35 & 48.45 \\
& Hydra         & 33.70 & 44.33    & 40.28 & 42.54 \\
& \ourscell{\approach} & \ourscell{59.89} & \ourscell{62.79} & \ourscell{49.01} & \ourscell{52.11} \\
\midrule

\multirow{9}{*}{\textbf{MiMo-V2.5}}
& Target-Local  & 40.27 & 44.88 & 37.18 & 38.87 \\
& RAG           & 48.55 & 49.15 & 43.10 & 43.10 \\
& LLMLingua2    & 29.26 & 35.95 & 29.58 & 34.37 \\
& LLM Summarize & 28.22 & 31.18 & 31.83 & 31.27 \\
& LongCodeZip   & 42.85 & 46.03 & 43.10 & 46.20 \\
& SWE-Pruner    & 52.82 & 54.36 & 40.56 & 41.69 \\
& InlineCoder   & 41.86 & 49.15 & 42.25 & 48.17 \\
& Hydra         & 35.01    & 40.55    & 34.93 & 40.56 \\
& \ourscell{\approach} & \ourscell{57.86} & \ourscell{59.84} & \ourscell{46.20} & \ourscell{50.42} \\

\bottomrule
\end{tabular}
}
\vspace{-3mm}
\end{table}

Table~\ref{tab:rq1_main_results} reports the Pass@1 results on DevEval and RepoExec under 4K and 8K context budgets.
\approach achieves the highest Pass@1 in all 16 settings.
On average, it improves over the strongest baseline by 5.15 and 4.36 percentage points on DevEval and by 3.73 and 4.65 percentage points on RepoExec under the 4K and 8K budgets, respectively.
The improvement ranges from 0.29 percentage points (GPT-5-mini on RepoExec at 4K, where LLM Summarize is the strongest baseline) to 10.14 percentage points (DeepSeek-V3.2 on RepoExec at 4K).
Paired bootstrap tests confirm that these gains are statistically significant: for each benchmark and budget, the 95\% confidence interval of the Pass@1 difference excludes zero.

Table~\ref{tab:rq1_main_results} also reveals why the baselines fall short.
First, the target location alone is not enough: Target-Local, which receives the ground-truth target location, trails \approach by 9.02--17.59 percentage points in every setting, indicating that much of the gain comes from evidence beyond the target file.
Second, shortening code as plain text discards evidence: LLMLingua2 and LLM Summarize fall below RAG in 30 of the 32 comparisons (the exceptions are LLM Summarize with GPT-5-mini on RepoExec), suggesting that token-level pruning and free-form rewriting remove identifiers and signatures that the generated code depends on, whereas LongCodeZip, which compresses at function granularity, narrows this gap and matches or exceeds RAG on RepoExec in all eight settings.
Third, no baseline is consistently second best: the strongest baseline alternates among InlineCoder, SWE-Pruner, LongCodeZip, LLM Summarize, and RAG across settings, whereas \approach ranks first in all 16 settings.

\begin{tcolorbox}[
  enhanced,
  colback=gray!15,
  colframe=gray!15,
  boxrule=0pt,
  arc=0pt,
  left=6pt,
  right=6pt,
  top=4pt,
  bottom=4pt,
  borderline west={2pt}{0pt}{gray!70}
]
\textbf{Answer to RQ1:}
\approach achieves the highest Pass@1 in all 16 settings, improving over the strongest baseline by 3.73--5.15 percentage points on average.
It outperforms Target-Local, which is given the ground-truth target location, by at least 9.02 percentage points, indicating that the evidence collected beyond the target file is a major source of the gain.
\end{tcolorbox}

\subsection{RQ2: Cost-Effectiveness Analysis}

RQ2 examines whether \approach provides a more cost-effective way to improve repository-level code generation than simply increasing the context budget or using a stronger backbone model.
We study two scaling strategies.
\textbf{Context-budget scaling} increases the context budget for the strongest non-\approach baseline, allowing the agent to receive more repository text under the same backbone model.
\textbf{Backbone-model scaling} replaces the backbone with a stronger model under the standard Mini-SWE-Agent scaffold.
For cost, we report the monetary cost of all recorded LLM calls, including prompt and generation tokens.
Repository parsing and graph construction in \approach are deterministic and involve no LLM calls, so they add no inference cost.

\begin{figure}[t]
\centering
\includegraphics[width=1.0\columnwidth]{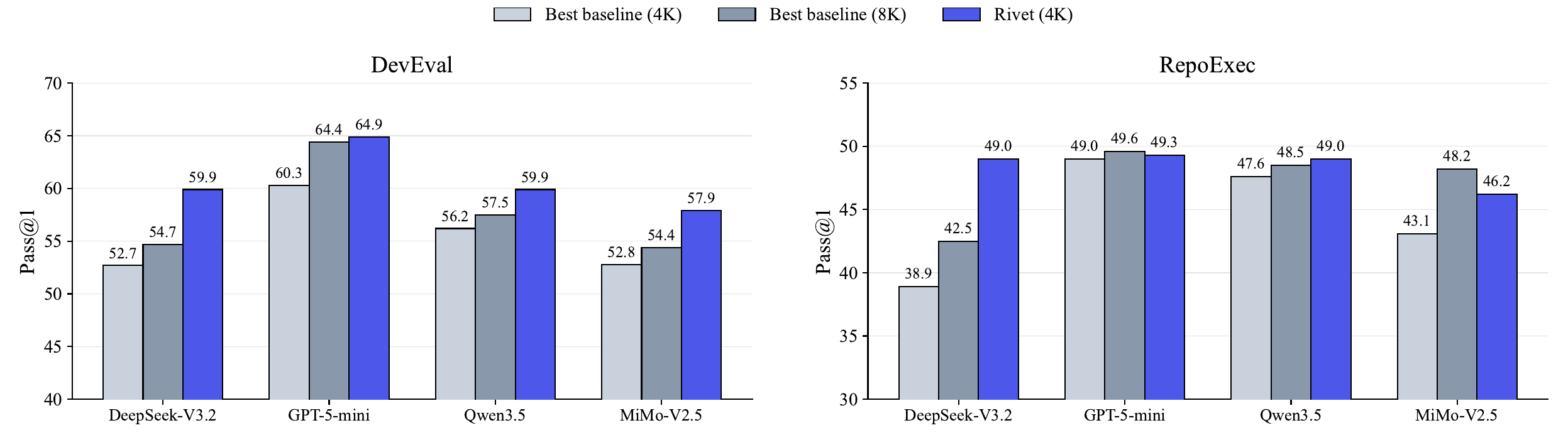}
\caption{Context-budget scaling. For each setting, we compare the strongest non-\approach baseline at the 4K and 8K budgets with \approach at the 4K budget. With twice the budget, the strongest baseline exceeds \approach in only two of the eight settings, both on RepoExec.}
\label{fig:rq2_budget}
\vspace{-2mm}
\end{figure}

\textbf{Context-budget scaling.}
This analysis reuses the RQ1 results and asks whether giving the strongest non-\approach baseline twice the context budget can match \approach.
For each backbone model and benchmark, we compare the strongest non-\approach baseline under the 4K and 8K budgets with \approach under the 4K budget, as shown in Figure~\ref{fig:rq2_budget}.
Averaged over the eight model--benchmark settings, doubling the budget raises the strongest baseline from 50.07 to 52.47 Pass@1, which is still below the 54.51 that \approach achieves with the 4K budget.
On DevEval, \approach under the 4K budget outperforms the strongest 8K baseline for all four backbone models.
On RepoExec, \approach under the 4K budget outperforms the strongest 8K baseline for DeepSeek-V3.2 and Qwen3.5-397B-A17B, and trails it by only 0.28 and 1.97 percentage points for GPT-5-mini and MiMo-V2.5, respectively.
These results suggest that, under these budgets, which evidence enters the context matters more than how much context the agent receives.

\textbf{Backbone-model scaling.}
This analysis examines whether using \approach with a cheaper backbone can provide a better accuracy--cost tradeoff than switching to a stronger backbone under the standard Mini-SWE-Agent scaffold.
We compare two representative pairs: DeepSeek-V3.2 + \approach versus DeepSeek-V4-Pro + Mini-SWE-Agent, and GPT-5-mini + \approach versus GPT-5 + Mini-SWE-Agent.
For each pair, Table~\ref{tab:rq2_model_scaling} reports Pass@1, measured LLM inference cost, and inference cost per passed task.
Table~\ref{tab:rq2_model_scaling} shows that \approach on cheaper backbones achieves a better Pass@1--inference-cost tradeoff than using stronger backbones with the standard Mini-SWE-Agent scaffold.
For the DeepSeek pair, DeepSeek-V3.2 + \approach improves Pass@1 by 8.71 percentage points on DevEval and 4.78 percentage points on RepoExec, while keeping inference cost comparable and reducing the cost per passed task.
For the GPT pair, GPT-5-mini + \approach outperforms GPT-5 + Mini-SWE-Agent by 6.03 percentage points on DevEval and 1.98 percentage points on RepoExec, while reducing inference cost by 73.3\% and 84.5\%, respectively.
In both pairs, improving the repository context is a more cost-effective way to raise Pass@1 than upgrading the backbone.

\begin{table}[t]
\centering
\caption{Backbone-model scaling. Infer. Cost is the total LLM inference cost (USD) over all tasks of a benchmark, and Cost/Pass divides it by the number of passed tasks. \approach uses the 4K budget.}
\label{tab:rq2_model_scaling}

\setlength{\tabcolsep}{2.5pt}
\renewcommand{\arraystretch}{1.08}
\begin{tabularx}{0.9\columnwidth}{@{}
l
>{\raggedright\arraybackslash}X
c
c
c
@{}}
\toprule
\textbf{Dataset} &
\textbf{Setting} &
\textbf{Pass@1} &
\textbf{Infer. Cost} &
\textbf{Cost/Pass} \\
\midrule

DevEval
& DeepSeek-V3.2 + \approach
& 59.89
& \$1.4473
& \$0.0013 \\
& DeepSeek-V4-Pro + Mini-SWE-Agent
& 51.18
& \$1.3936
& \$0.0015 \\

RepoExec
& DeepSeek-V3.2 + \approach
& 49.01
& \$0.2462
& \$0.0014 \\
& DeepSeek-V4-Pro + Mini-SWE-Agent
& 44.23
& \$0.2493
& \$0.0016 \\

\midrule

DevEval
& GPT-5-mini + \approach
& 64.93
& \$2.0799
& \$0.0018 \\
& GPT-5 + Mini-SWE-Agent
& 58.90
& \$7.7889
& \$0.0072 \\

RepoExec
& GPT-5-mini + \approach
& 49.30
& \$0.2095
& \$0.0012 \\
& GPT-5 + Mini-SWE-Agent
& 47.32
& \$1.3527
& \$0.0081 \\

\bottomrule
\end{tabularx}

\end{table}

\begin{tcolorbox}[
  enhanced,
  colback=gray!15,
  colframe=gray!15,
  boxrule=0pt,
  arc=0pt,
  left=6pt,
  right=6pt,
  top=4pt,
  bottom=4pt,
  borderline west={2pt}{0pt}{gray!70}
]
\textbf{Answer to RQ2:}
With a 4K budget, \approach outperforms the strongest baseline given an 8K budget in six of the eight settings (54.51 vs.\ 52.47 Pass@1 on average).
GPT-5-mini with \approach outperforms GPT-5 while reducing inference cost by 73.3\%--84.5\%, and DeepSeek-V3.2 with \approach outperforms DeepSeek-V4-Pro at comparable cost.
\end{tcolorbox}

\subsection{RQ3: Ablation Study}

To assess the contribution of each stage, we construct three controlled ablated variants of \approach.
Each variant disables one stage while keeping the remaining stages and experimental settings unchanged.
\textbf{1) w/o Signal Grounding} replaces task-aware grounding with similarity-based retrieval, which selects snippets only by their similarity to the task description.
\textbf{2) w/o Utility Linking} keeps the grounded snippets but removes relation-based expansion, so it does not follow imports, symbol references, caller usages, or configuration relations to collect repository evidence.
\textbf{3) w/o Budget Shaping} keeps grounding and linking but removes tiered packaging and representation shaping, greedily adding evidence in relevance order until the token budget is exhausted.

\begin{table}[t]
\centering
\caption{Ablation study under the 4K context budget.}
\label{tab:rq3_ablation}
\setlength{\tabcolsep}{3pt}
\renewcommand{\arraystretch}{1.08}

\begin{tabularx}{0.9\columnwidth}{@{}
>{\centering\arraybackslash}p{3.5cm}
>{\raggedright\arraybackslash}X
>{\centering\arraybackslash}p{1.7cm}
>{\centering\arraybackslash}p{1.7cm}
@{}}
\toprule
\textbf{Model} & \textbf{Variant} & \textbf{DevEval} & \textbf{RepoExec} \\
\midrule

\multirow{4}{*}{\textbf{GPT-5-mini}}
& w/o Signal Grounding & 43.89 & 30.70 \\
& w/o Utility Linking  & 48.05 & 35.49 \\
& w/o Budget Shaping   & 63.78 & 44.79 \\
& Full \approach       & \textbf{64.93} & \textbf{49.30} \\

\midrule

\multirow{4}{*}{\textbf{DeepSeek-V3.2}}
& w/o Signal Grounding & 31.01 & 25.63 \\
& w/o Utility Linking  & 38.36 & 29.58 \\
& w/o Budget Shaping   & 55.62 & 40.56 \\
& Full \approach       & \textbf{59.89} & \textbf{49.01} \\

\midrule

\multirow{4}{*}{\textbf{Qwen3.5-397B-A17B}}
& w/o Signal Grounding & 38.08 & 30.14 \\
& w/o Utility Linking  & 42.58 & 35.77 \\
& w/o Budget Shaping   & 58.68 & 47.89 \\
& Full \approach       & \textbf{59.89} & \textbf{49.01} \\

\midrule

\multirow{4}{*}{\textbf{MiMo-V2.5}}
& w/o Signal Grounding & 34.14 & 28.73 \\
& w/o Utility Linking  & 38.58 & 32.96 \\
& w/o Budget Shaping   & 56.93 & 44.79 \\
& Full \approach       & \textbf{57.86} & \textbf{46.20} \\

\bottomrule
\end{tabularx}

\end{table}

Table~\ref{tab:rq3_ablation} reports the ablation results under the 4K context budget.
Across all backbone models and datasets, \approach consistently outperforms the three ablated variants.
Removing Signal Grounding causes the largest performance drop, with an average Pass@1 decrease of 21.72 percentage points.
This variant even falls below RAG in all eight settings (e.g., 31.01 vs.\ 49.64 for DeepSeek-V3.2 on DevEval).
This suggests that the later stages amplify, rather than correct, grounding errors: when the anchors are wrong, Utility Linking expands from the wrong locations and fills the budget with evidence connected to irrelevant code.
Removing Utility Linking also substantially reduces performance, with an average Pass@1 decrease of 16.84 percentage points.
Without Utility Linking, the packed context contains the anchor files but none of the dependency, usage, and configuration evidence linked from them.
That this variant falls below even Target-Local in all eight settings indicates that locating the relevant files is not sufficient by itself; the gain of \approach comes from the evidence collected from them.
Removing Budget Shaping leads to a smaller but consistent average Pass@1 decrease of 2.88 percentage points (0.93--8.45 across settings).
This result suggests that tiered packaging and representation shaping further help preserve useful repository evidence under the token budget.

To further investigate the contribution of different repository evidence types, we conduct a fine-grained ablation study of Utility Linking on DevEval and RepoExec under a representative experimental setting, using DeepSeek-V3.2 with a 4K context budget.
Specifically, we construct three variants by independently disabling dependency, usage, and configuration relation-based expansion, respectively. All variants share the same task anchors produced by the complete Signal Grounding stage, while keeping the remaining components and experimental settings unchanged. This design isolates the contribution of each relation type within Utility Linking without introducing changes in anchor localization.
Table~\ref{tab:evidence_ablation} presents the results.
On RepoExec, disabling dependency-, usage-, or configuration-based expansion reduces Pass@1 by 8.16, 10.14, and 8.16 percentage points, respectively, with usage evidence showing the largest effect.
On DevEval, the corresponding drops are small (0.27, 0.38, and 0.49 percentage points, i.e., 5 to 9 of the 1,825 tasks), although removing all relations (w/o Utility Linking) costs 21.53 percentage points.
This contrast suggests that on DevEval the three relation types largely reach the same evidence, so the remaining types compensate for a disabled one, whereas on RepoExec each type contributes evidence that the others do not cover.

\begin{table}[h]
\centering
\caption{Fine-grained ablation of Utility Linking on DeepSeek-V3.2 under the 4K context budget.}
\label{tab:evidence_ablation}

\setlength{\tabcolsep}{6pt}
\renewcommand{\arraystretch}{1.08}

\begin{tabularx}{0.7\columnwidth}{@{}
>{\raggedright\arraybackslash}X
>{\centering\arraybackslash}p{1.7cm}
>{\centering\arraybackslash}p{1.7cm}
@{}}
\toprule
\textbf{Variant} & \textbf{DevEval} & \textbf{RepoExec} \\
\midrule

w/o Dependency    & 59.62 & 40.85 \\
w/o Usage         & 59.51 & 38.87 \\
w/o Configuration & 59.40 & 40.85 \\

\midrule

Full \approach    & \textbf{59.89} & \textbf{49.01} \\

\bottomrule
\end{tabularx}

\end{table}

\begin{tcolorbox}[
  enhanced,
  colback=gray!15,
  colframe=gray!15,
  boxrule=0pt,
  arc=0pt,
  left=6pt,
  right=6pt,
  top=4pt,
  bottom=4pt,
  borderline west={2pt}{0pt}{gray!70}
]
\textbf{Answer to RQ3:}
All three stages contribute to \approach.
Removing Signal Grounding or Utility Linking reduces Pass@1 by 21.72 and 16.84 percentage points on average, and Budget Shaping adds a smaller but consistent 2.88 percentage points.
Within Utility Linking, the three relation types are complementary on RepoExec, where removing any one of them costs 8.16--10.14 percentage points, but largely overlap on DevEval.
\end{tcolorbox}

\section{Discussion}
\subsection{How Does \approach Preserve Repository Evidence?}

\begin{figure*}[ht]
\centering
\includegraphics[width=\textwidth]{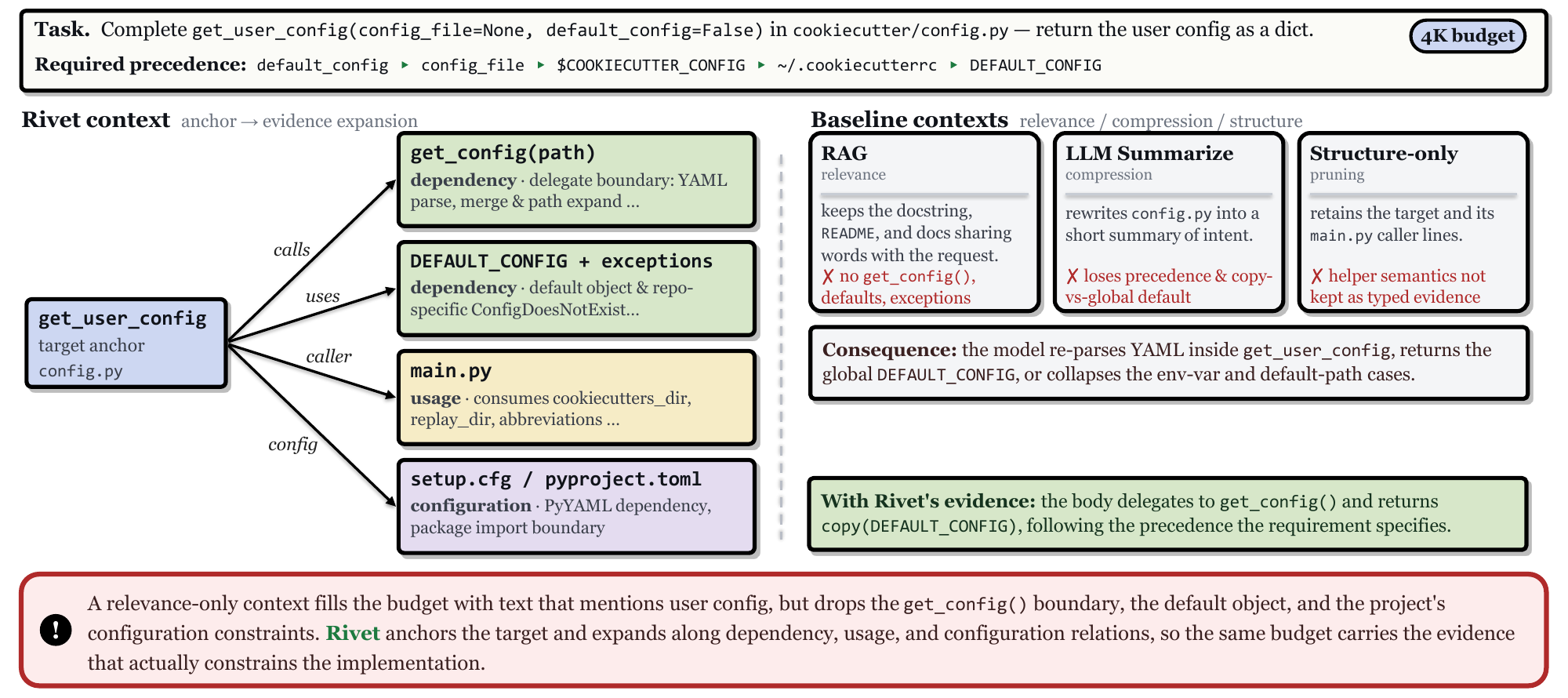}
\caption{Qualitative case for the RepoExec task \texttt{cookiecutter::get\_user\_config} under a 4K budget, contrasting the repository evidence retained by \approach with that of relevance, compression, and structure-only baselines.}
\vspace{-8pt}
\label{fig:case}

\end{figure*}

To show where the gain of \approach comes from, Figure~\ref{fig:case} walks through the RepoExec task \path{cookiecutter::get\_user\_config} under the 4K budget.
The requirement specifies a configuration-precedence rule, but the repository evidence needed to implement this rule is distributed across helper functions, default configuration objects, caller usages, and project configuration files.

Signal Grounding first anchors the task at \texttt{cookiecutter/config.py}, the file that contains \texttt{get\_user\_config}.
From this anchor, Utility Linking follows a call relation to the \texttt{get\_config()} helper, which parses, merges, and expands configuration values, even though the requirement never mentions it.
It further collects the \texttt{DEFAULT\_CONFIG} object and the repository-specific \texttt{ConfigDoesNotExist} exception through symbol references, the caller in \texttt{main.py} through usage relations, and the PyYAML dependency declared in \texttt{setup.cfg} and \texttt{pyproject.toml} through configuration relations.
The baseline contexts each miss a different part of this evidence.
RAG fills the budget with the docstring, README, and documentation that share words with the request, but includes neither \texttt{get\_config()} nor the default object and exceptions.
LLM Summarize rewrites \texttt{config.py} into a short summary of intent and loses the precedence rule and the distinction between returning a copy and returning the global default.
InlineCoder keeps the target and its caller lines in \texttt{main.py}, but not the helper semantics.
Budget Shaping then fits all four kinds of evidence into the 4K budget by rendering lower-priority evidence, such as the caller and configuration files, as signatures or summaries instead of full snippets.
With this context, the generated body delegates to \texttt{get\_config()} and returns \texttt{copy(DEFAULT\_CONFIG)}, following the precedence that the requirement specifies.
With the baseline contexts, the model instead re-parses YAML inside \texttt{get\_user\_config}, returns the global \texttt{DEFAULT\_CONFIG}, or collapses the environment-variable and default-path cases.
This case reflects the pattern in RQ1: the baseline contexts are relevant to the requirement, yet the evidence that decides how to implement it lies in code that the requirement never mentions.

\subsection{How Do Graph Components Affect \approach's Performance?}

\begin{table}[t]
\centering

\caption{Contribution of graph-based components on DeepSeek-V3.2 under the 4K budget.}
\vspace{-8pt}
\label{tab:graph_ablation}

\setlength{\tabcolsep}{3pt}
\renewcommand{\arraystretch}{1.08}

\begin{tabularx}{0.8\columnwidth}{@{}
>{\raggedright\arraybackslash}X
>{\centering\arraybackslash}p{2.4cm}
>{\centering\arraybackslash}p{2.4cm}
@{}}
\toprule
\textbf{Variant} & \textbf{DevEval} & \textbf{RepoExec} \\
\midrule
w/o Graph Ranking         & 59.62        & 41.41 \\
w/o Structural Grounding  & 59.56        & 40.00 \\
Dependency Edges Only     & 59.45        & 40.28 \\
Full \approach            & \textbf{59.89} & \textbf{49.01} \\
\bottomrule
\end{tabularx}

\end{table}

Both the structural grounding signal and Utility Linking rely on statically extracted repository relations, which may miss dynamic behavior such as runtime imports or reflection~\cite{bouzenia2024dypybench}.
\approach does not rely on these relations alone: Signal Grounding fuses the structural signal with lexical and semantic signals, so missing edges can reduce the evidence that Utility Linking collects but cannot by themselves remove a file from the anchors.

To examine this, we construct three variants on DeepSeek-V3.2 under the 4K budget.
\textbf{w/o Graph Ranking} keeps evidence collection but removes the propagated relevance from Budget Shaping, ordering evidence by grounding relevance and tier alone.
\textbf{w/o Structural Grounding} disables the structural channel in Signal Grounding while keeping the lexical and semantic signals.
\textbf{Dependency Edges Only} restricts Utility Linking to import and symbol-reference relations, removing usage and configuration relations.

As shown in Table~\ref{tab:graph_ablation}, the graph components matter little on DevEval, where each variant loses at most 0.44 percentage points (8 of the 1,825 tasks), but substantially on RepoExec, where the drops range from 7.60 to 9.01 percentage points and removing the structural grounding signal has the largest effect.
This pattern mirrors the fine-grained ablation in RQ3 and suggests that on RepoExec the graph contributes evidence that text similarity does not recover; restricting Utility Linking to import and symbol-reference relations alone already costs 8.73 percentage points.

This ablation measures how much the graph contributes, not how robust \approach is to wrong or missing edges; we leave a controlled study of edge noise to future work.

\subsection{How Accurately Does Signal Grounding Localize Target Files?}

Signal Grounding provides the starting anchors for subsequent evidence expansion. To directly assess its localization quality, we measure whether the ground-truth target file appears among the top-ranked anchor files. The ground-truth target file is used only for this analysis and never for context construction. We report Target@$k$, defined as the percentage of tasks whose target file appears within the top-$k$ ranked files, together with mean reciprocal rank (MRR), which reflects the overall ranking position of the target file.

Table~\ref{tab:target_localization} reports the results. On DevEval, Signal Grounding identifies the target file at rank 1 for 90.36\% of the tasks and within the top 5 for 97.97\%, with an MRR of 0.939. The corresponding results on RepoExec are 98.59\%, 99.44\%, and 0.989, respectively. These results show that Signal Grounding provides reliable starting points for Utility Linking. This accuracy is partly expected, since benchmark requirements often name the target function explicitly and \approach exploits such mentions (Section~\ref{sec:overview}). Localization alone, however, does not explain the gains of \approach: Target-Local, which is given the ground-truth location, still trails \approach by 9.02--17.59 percentage points (RQ1).

\begin{table}[h]
\centering
\caption{File-level target localization performance of Signal Grounding.}
\vspace{-8pt}
\label{tab:target_localization}

\setlength{\tabcolsep}{5pt}
\renewcommand{\arraystretch}{1.08}

\begin{tabular}{@{}lcccc@{}}
\toprule
\textbf{Dataset} & \textbf{\# Tasks} & \textbf{Target@1} & \textbf{Target@5} & \textbf{MRR} \\
\midrule
DevEval  & 1,825 & 90.36\% & 97.97\% & 0.939 \\
RepoExec & 355   & 98.59\% & 99.44\% & 0.989 \\
\bottomrule
\end{tabular}
\end{table}

\section{Threats to Validity}

\textbf{Generalizability.}
Our experiments focus on Python repositories, and our implementation relies on Python-specific parsing, import resolution, and execution environments.
Only these front-end components are language-specific: the grounding signals, graph propagation, and budget packing operate on files, snippets, and typed relations, and apply to other languages once a parser extracts these relations.
Validating \approach on other languages is left to future work.

\textbf{Task representativeness.}
DevEval and RepoExec provide executable and reproducible evaluation, but both focus on function-level generation within existing repositories.
They do not cover multi-file changes such as feature addition and bug fixing, where locating the code to change is itself part of the task.
Future work should further evaluate \approach on broader benchmarks for issue resolution and feature evolution, such as SWE-Bench~\cite{jimenez2024swe}, FEA-Bench~\cite{li2025fea}, and FeatBench~\cite{chen2025featbench}.

\textbf{Comparison fairness.}
The compared methods were designed for different context-construction settings, including retrieval, compression, and structure-aware selection.
To reduce confounding factors, all methods use the same task requirement, repository files, backbone model, prompt template, context budget, decoding strategy, agent scaffold, and execution environment.
This protocol makes context construction the main controlled variable, although method-specific tuning could further improve individual baselines.

\textbf{Hyperparameters.}
\approach has several hyperparameters (Section~\ref{sec:implementation_details}), which we keep fixed across all benchmarks, models, and budgets rather than tuning them per setting; we do not study their sensitivity, so other values may change the absolute results.

\textbf{Metric validity and data leakage.}
We use Pass@1 on official benchmark tests as the primary metric because it directly measures executable task success. However, passing tests does not guarantee full semantic correctness. Public benchmarks may also have appeared in the pretraining data of large language models. This risk may affect absolute scores, but all methods are evaluated with the same backbone models and protocol, which reduces its impact on relative comparisons. During context construction, we use only task requirements and repository files; reference implementations, reference dependencies, benchmark tests, and the ground-truth target location are used only for evaluation and analysis.

\section{Related Work}

\textbf{Retrieval and Compression for Code Context.}
A line of work controls repository context by selecting or shortening code before prompting LLMs.
Retrieval-based methods select files, functions, or code snippets according to lexical or semantic similarity to the task requirement~\cite{lewis2020retrieval,liu2024repobench,parvez2021retrieval,zhang2023repocoder}.
Code search tools retrieve code by textual or structural similarity to a query~\cite{kim2018facoy,luan2019aroma}, and repository-level completion methods adapt such retrieval to cross-file context, for example by iterating retrieval with generation~\cite{zhang2023repocoder}, deciding when to retrieve~\cite{wu2024repoformer}, learning what to retrieve~\cite{wang2025rlcoder}, or following data flow~\cite{cheng2024dataflow}.
Prompt compression methods reduce token usage by pruning low-importance tokens or rewriting retrieved repository content~\cite{pan2024llmlingua,jiang2023llmlingua}.
Code-specific compression methods further exploit code granularity by selecting functions or blocks under a context budget~\cite{shi2025longcodezip}.
These methods make repository context more relevant or compact, but they mainly optimize task similarity or context length.
As a result, the selected context may still lack the repository evidence that the implementation depends on~\cite{gu2025retrieve}.
Moreover, LLMs do not use long contexts uniformly~\cite{liu2024lost}, so a longer context does not necessarily carry more useful evidence.
\approach also selects and compresses context, but it decides what to keep by following repository relations from the task location rather than by similarity to the requirement, and it compresses by lowering the representation level of each snippet rather than by pruning tokens.

\textbf{Structure-aware Repository Context Construction.}
Another line of work uses repository structure or agent states to guide context construction.
Structure-aware methods, such as InlineCoder and Hydra, exploit repository relations to retrieve or inline context connected to the task~\cite{hu2026line,le2026not,liu2024graphcoder,phan2025repohyper}.
These methods build on a long line of work that represents programs as graphs of data and control dependencies or calls~\cite{allamanis2017learning,nguyen2019graph,salis2021pycg}.
Other methods refine the context during generation, by pruning agent observations according to the current coding goal, as in SWE-Pruner~\cite{wang2026swe}, or by iteratively retrieving project context with compiler feedback~\cite{bi2024iterative}.
These methods improve over pure text-based retrieval by incorporating structural or interaction signals.
\approach differs from them in three respects.
First, it collects dependency, usage, and configuration evidence from the same anchors, whereas InlineCoder inlines callers and callees and Hydra retrieves dependencies and usage examples.
Second, it treats the budget as a packing problem and lowers the representation level of each snippet before dropping it, rather than truncating or pruning context.
Third, it constructs the context deterministically without LLM calls, whereas InlineCoder requires a draft generation and SWE-Pruner relies on a learned skimmer.

\section{Conclusion}

In this paper, we characterized context utility dilution, a failure mode in which the context given to a repository-level coding agent appears relevant to the task but lacks the repository evidence that the implementation depends on.
Our key observation is that much of this evidence is reachable from the code that the task touches through static repository relations.
Based on this observation, we proposed \approach, which locates task-related code anchors, follows dependency, usage, and configuration relations from them to collect evidence, and packs this evidence under the budget with a representation ladder.
On DevEval and RepoExec with four backbone models, \approach achieves the highest Pass@1 in all 16 settings and improves over the strongest baseline by 3.73--5.15 percentage points on average.
With a 4K budget, it outperforms the strongest baseline given an 8K budget in six of the eight settings, and it enables GPT-5-mini to outperform GPT-5 at 73.3\%--84.5\% lower inference cost.
These results suggest that, under constrained budgets, which evidence enters the context matters more than how much context the agent receives.
In future work, we plan to extend \approach to other programming languages and to multi-file tasks such as issue resolution and feature implementation.


\bibliographystyle{ACM-Reference-Format}
\bibliography{references}

\end{document}